\documentclass[journal]{IEEEtran}
\usepackage{cite}

\ifCLASSINFOpdf
   \usepackage[pdftex]{graphicx}
   \graphicspath{{../pdf/}{../jpeg/}}
   \DeclareGraphicsExtensions{.pdf,.jpeg,.png}
\else
   \usepackage[dvips]{graphicx}
   \graphicspath{{../eps/}}
   \DeclareGraphicsExtensions{.eps}
\fi
\usepackage{hyperref}
\usepackage{flushend}

\begin{document}
%
\title{A Conceptual Marketplace Model for IoT Generated Personal Data}
%
%
%

\author{Victor Molina, Marta Kersten-Oertel, Tristan Glatard\\
Department of Computer Science and Software Engineering\\
Concordia University, Montreal, Canada}

\maketitle

\begin{abstract}
We propose a decentralized conceptual marketplace model for IoT generated personal data. Our model is based on a thorough analysis of personal data in a marketplace context, with specific focus on the challenges presented by commercializing IoT generated personal data. Our model introduces a novel perspective on the commercialization of personal data for a marketplace context via risk evaluation and a data licensing framework. We have designed our model to be centered around protecting the privacy and data rights of data generators through model components that effectively assess and modify transaction risks, and formalize transaction agreements by establishing rights of data use and access between buyer and seller. Our model could serve as a blueprint  to  inform  the  implementation  of  a personal data marketplace that respects  privacy and  ownership.  
\end{abstract}


%
\IEEEpeerreviewmaketitle

\section{Introduction}
%
%
%
%
\IEEEPARstart{C}{onnected} devices have seen an exponential rise in popularity in recent years and this trend is not expected to wane. It is, therefore, crucial that we consider how the field of privacy research will be affected by this technological trend. Individuals are likely to generate personal data through the use of connected devices at an unprecedented rate, which has repercussions for privacy and data ownership as we know it. Ideally, any new technology should empower users and grant them new avenues to exercise their digital rights, including furthering their involvement in the way their personal data is used. As users generate more personal data than ever before through their devices, we must seriously consider what ramifications this will have for the way personal data is used. Based on the assumption that the sale of personal data by users is an inevitable next step in the growth of the Internet of Things (IoT), and an efficient way to spread the wealth created from digital technologies, this paper focuses on the specific question: \emph{What is the best model to use when dealing with the sale of personal data while keeping privacy intact? }

In 2011, a World Economic Forum (WEF) initiative resulted in a report on personal data and its uses shedding light on a vision of oncoming shifts regarding the commercialization of personal data in the near future. The report claims that personal data, what it refers to as ``digital data created by and about people\cite{PersonalData:wec}", is gaining importance and salience as it generates ``a new wave of opportunity for economic and societal value creation\cite{PersonalData:wec}". Indeed, this new dawn for personal data stems from its value as a new-asset class in a post-industrial context\cite{PersonalData:wec}.  The main idea propelling this vision is the concept that industries will find personal data to be the fuel they need to innovate their products and services. Beyond its potential uses for industry, personal data also stands to become a valuable asset for data generators themselves, who can potentially reap benefits from a new ecosystem that is flourishing around the uses of personal data. Personal data has been referred to by Maglena Kuneva, the European Consumer Commissioner, as ``the new oil of the Internet and the new currency of the digital world\cite{PersonalData:wec}". At the same time, the WEF report highlights how fragmented the state of the personal data ecosystem is. It is clear that there are concerns for privacy, as well as others, that are glaringly obvious. As the rally behind the WEF's vision grows and personal data continues to be hailed as a transformative resource in modern economies, it will become critical to address these concerns in full\cite{PersonalData:wec}. The report states that current institutions and frameworks ``fall short of providing the legal and technical infrastructure needed to support a well-functioning digital economy\cite{PersonalData:wec}". There are risks associated with the use of personal data for those who generate it and their privacy. IoT privacy risks represent a specific kind of privacy threat given the personal and intimate nature of the data that IoT devices can generate. There is also the additional risks stemming from the types of insights and discoveries that can be achieved by processing IoT generated personal data through data mining procedures and other metadata generating techniques. The kind of insight that can result from such undertakings, using IoT generated personal data, can reveal intimate and personal dimensions of the data generator's private life, thereby exposing them to serious hazards.

Our approach to providing a transaction model for IoT generated personal data is divided into two parts: our analysis of the subject followed by our proposed marketplace model. Our methodology is to move from the abstract concept of personal data, to personal data as a marketplace product, and then to propose a marketplace model in which such a product could be effectively bought and sold without compromising the privacy of the data subject. This shift from abstract to practical requires that we look at the concept of personal data in the abstract to find the qualities that will become challenging in the pursuit of bringing personal data into a marketplace context. We will perform an analysis of these challenges and design model components to address them. Then, we will provide a description of the full marketplace model, its components, and its interactions. Our proposed response is a model that provides the functionality required by our original question: allowing individuals to buy and sell personal data, protecting personal data rights, and protecting the privacy of data generators.

The starting point for this analysis is a discussion of the context in which personal data can be transacted: namely the personal data marketplace. Our analysis  involves a discussion of the challenges faced by marketplaces in general, as well as a focus on the specific challenges that a personal data marketplace would present. This kind of exploration calls for a definition of personal data in a marketplace context. This definition includes a description of how ownership is established and understood for personal data. This description must highlight the connection between data and the user that generates it, as well as how this constitutes ownership, or ``the right to" use, dispose, or sell such data. Our definition of personal data in a marketplace context must also describe the basis on which we can say personal data holds any marketplace value at all: it should take into account the elements that make personal data a desirable asset in a marketplace context. As we develop this analysis of personal data in a marketplace context, we will also develop specific model components that address any challenges we encounter to the commercialization of personal data. 

The second part of our approach is the model proposal that connects the model components stemming from our analysis, and also provides the functionality outlined in our original research question. This proposed model must include a clear description of how the marketplace operates, who are its actors, and how data transactions occur between the seller and buyer. The description of the transaction process must incorporate marketplace mechanisms that will safeguard the marketplace from the challenges generated by the commercialization of personal data, as well as the different marketplace interactions that enable successful transactions between buyers and sellers. Essentially, our proposed marketplace model will describe a set of components, mechanisms, and interactions, that provide the expected functionality of a personal data marketplace and allow the marketplace to succeed in its goal of protecting the privacy and rights of the data subject. 

The model we describe in this paper is a conceptual one. It is not meant to be a recommendation of a specific system, or technical implementation for a personal data marketplace. It will include technical recommendations, but it is primarily meant as a blueprint to inform the implementation of a personal data marketplace that respects a set of values: privacy and ownership. After all, technology will change, advance, and innovate in ways we cannot foresee, but the need to protect personal data rights and the privacy of data generators is something that we consider to be a constant, unchanging, factor in the discussion of the commercialization of personal data.

\section{Personal Data Challenges \& Related Model Components}
\subsection{The Marketplace Context}
This paper aims to describe an appropriate model through which the parties involved in the transacting of personal data can benefit. As a starting point for this model, we propose a data marketplace as the core of a framework which facilitates the buying and selling of personal data while addressing the risks involved. The reason for choosing a marketplace as a base involves some insight from marketplace theory: a broad discussion on how marketplaces enable transactions between parties follows ahead to further clarify this decision. 

The transaction between a \emph{data generator} and a \emph{data buyer}, much like any transaction between buyer and seller, must begin somewhere. Regardless of which side you are facing from, the buyer or the seller, the transaction begins with a search and its associated costs. The \emph{search cost} is the price that either a buyer or a seller must pay in order to find someone to willing to enter into a transaction with them. Therefore, the price tag of a given item involves more than just the valuation of the item itself, it involves the search costs associated with locating the necessary party that would allow the transaction to take place\cite{ebayIdeas:dushnitsky}. Search costs may be particularly onerous when it comes to the buying and selling of personal data given its extremely flexible nature. The categorical limitations on personal data are vague at best, personal data can be generated in many ways and therefore take many forms. From riding a bike to work every day to the percentages of enzymes in the body at a given moment, any of these could form a data set that a data generator might be interested in selling.

The potential specificity of such data sets means that finding a party interested in transacting could be costly, in terms of time or other resources. As noted by the WEC, the things that personal data can be used for are also largely varied. As an example, a shoe company may be interested in data from suburban middle aged women that actively jog in order to optimize the design of a new sneaker line and hit sale targets within a specific demographic. To depict the level of specificity, take for example an ice cream company that may be interested in data sets comprised of the dietary logs of people from a specific age range in order to determine the average ice cream consumption of a given populations and change their pint size accordingly. Finding individuals that have these specific data sets for sale would be costly. A marketplace provides a solution to this challenge, it is an environment in which buyers and sellers can easily connect regardless of specificity, therefore driving down search costs considerably. The strength of the marketplace is its matching ability, it ``accommodates those who own and those who seek\cite{ebayIdeas:dushnitsky}". 
	
By driving down search costs, a data marketplace would help facilitate transactions between those who want to sell their data, and those looking to buy. Unfortunately, searching is not the only obstacle to buying and selling data. Marketplaces are defined categorically by the type of assets that are traded in them. As defined by Dushnitsky and Klueter, knowledge marketplaces are those where knowledge owners and knowledge seekers meet. The authors describe knowledge owners as  ``those who arrive at a market with knowledge assets\cite{ebayIdeas:dushnitsky}", and knowledge seekers as ``those who arrive at a market with the intention to transact with knowledge owners\cite{ebayIdeas:dushnitsky}". These definitions are not at all different from our own descriptions of the data generator and the data buyer. The value of personal data is not inherently found in its physical form, but instead it is found in the insight that personal data can provide: knowledge that can be utilized to optimize, enhance, advance and innovate. Conceptually, there is no distinction between data, and what Dushnitsky and Klueter refer to as ``knowledge assets\cite{ebayIdeas:dushnitsky}". Knowledge assets are particularly inclined to cause two serious problems for buyers and sellers: the \emph{adverse selection problem} and the \emph{expropriation problem}. The adverse selection and expropriation problems refer to the specific risks data sellers or data buyers face in the marketplace. These two problems are so severe that they alone can be responsible for bringing about a marketplace breakdown.

The \emph{adverse selection problem} refers to an information asymmetry between buyer and seller. Evaluating the quality of a knowledge asset is difficult. Malicious actors can take advantage of this asymmetry to pose as reputable data sellers and effectively deceive buyers into purchasing junk data\cite{ebayIdeas:dushnitsky}. An element of trust is required to engage in such a transaction, and unfortunately when profit is on the line, those who seek to take advantage of the adverse selection problem will likely never be in short supply. It is not hard to imagine how one would go about taking advantage of the adverse selection problem: all a malicious actor needs to do is buy a dataset of the data they plan on imitating, mimicking the specific data characteristics and distributions, and then falsely generating the data using unspecified means and back-logging it to disguise it with historical validity. This requires a marketplace mechanism that can somehow alleviate the information asymmetry, either via data validation or other means, and diffuse the adverse selection problem, or at least dilute it to such a level that the risk is so minimal it is easily tolerable by transacting parties. 

On the other hand, the \emph{expropriation problem} is one that puts the seller at risk. This problem involves ownership: as Dushnitksy \& Klueter note, knowledge replication has very little cost associated with it, and sharing knowledge assets does not deplete them\cite{ebayIdeas:dushnitsky}. This goes double for personal data, replicating and sharing it is only a click away, at virtually no cost. Once a seller dispenses with their data to a buying party, that entity can then resell the data for pure profit, a chain that could go on endlessly with second-hand, third-hand up to nth-hand sellers. This asymmetry must also be mitigated in order to increase the stability of the marketplace. Part of the value of personal data comes from its exclusivity, the other part comes from its uniqueness, therefore the resell problem theoretically devalues the data each time it is sold second hand. 

Data generators who have the intent of selling personal data face the expropriation problem and the adverse selection problem in tandem. To prove the legitimacy of their data, they must reveal some of it prior to sale or risk the cost of widening the information asymmetry. But if they reveal their data prior to sale they risk it being imitated, devaluing it or outright losing it. This calls for data handling protocols between buyer and seller that can overcome these two challenges in clever ways. Although classical knowledge assets include things like inventions and other intellectual properties, which are well regulated by now and entrenched in most modern legal systems, personal data is a nascent type of knowledge asset that does not enjoy the same type of institutional support and protection. For this reason, data marketplaces are especially vulnerable to the adverse selection, and expropriation problems.

\subsection{Data Ownership in a Marketplace Context}
In order to provide a model that describes the transacting of personal data, we must first clarify what we are referring to when we talk about personal data. From the name itself we gather that it is something that has two kinds of qualities: the personal qualities and the quantifiable qualities. Each of these two categories contributes towards the meaning of personal data, but the concept truly is more than the sum of its two parts, part personal and part data. The combination of the two allows for specific qualities that must be considered as well. The most simple and straightforward definition we can give of personal data stems from the acts and circumstances which instantiate it. Namely, we refer to the event in which a person decides to use a form of technology to transform some aspect of their personal experience into a quantifiable form that can be physically stored. In fact, this isn't too far off from the official definition of personal data used by the European Union's General Data Protection Regulation (GDPR), which defines personal data as ``any information relating to an identified or identifiable natural person ('data subject')\cite{GDPR:EU}".  What the GDPR definition is missing when compared to what we give above is the aspect of instantiation. The GDPR definition is a more general definition meant to address the numerous types of personal data that can exist, and how they vary depending on the way they are instantiated. A taxonomy of personal data has been proposed based on the varying degrees of proprietary entitlement that a person has to said data based on how it becomes instantiated\cite{taxonomy:malgieri}. 

In its initial form, the taxonomy was suggested with three different types of personal data: types where there is a strong proprietary relationship between a person and their data, types with an intermediate proprietary relationship, and types with a weak relationship\cite{taxonomy:malgieri}.  Data types with a strong relationship refer to those where a data subject directly and willingly provides information to a second party, an example of this is when a person fills in a form online, or willingly provides personal information on a public digital space. The relationship here is very clear, and it is easy to track the proprietary source of the data. It's obvious that without a person to provide it, this type of data is completely inaccessible\cite{taxonomy:malgieri}. Data types with an intermediate relationship refer to those were the data is provided indirectly, but via ``real" means of instantiation. An example of this is the data that is generated as a byproduct of using GPS or via the use of a web browser. The relationship here is still clear, although not as strong as the first type since the data subject is not required to provide the data directly, it can be a byproduct of their digital activity, yet it is still directly linked to the data subject\cite{taxonomy:malgieri}. Lastly, data types with a weak relationship are those where new data is created by applying some process to personal data. For example, forecasting or predictive information generated by analysis of personal data, which have a very weak proprietary link to the data subject themselves\cite{taxonomy:malgieri}.  The taxonomy essentially defines types of data based on how many deviations a data type takes away from the data subject, with personal metadata at the end of the chain being essentially data about personal data.

With this taxonomy in mind, we can see that personal data generated by the data subject prior to transacting has an equally strong proprietary relationship, if not stronger, than data willingly submitted on a digital platform. It is personal data that is still completely in control and under the ownership of the data subject. Here we make a distinction between personal data in general, and personal data as defined in the context of a data marketplace. Unless explicitly stated, from here on in when we refer to personal data we are talking about data generated and stored by the data subject with the intention of capitalizing on its value via a data marketplace transaction. What is critical is that at its inception personal data is completely within the proprietary jurisdiction of the data subject and generator. Without clarifying this aspect of our definition of personal data, a data marketplace model cannot exist, given it violates one of the fundamental tenets of the marketplace: how do you sell something you don't own? It is crucial to crystalize two aspects of personal data to place it in the context of a marketplace : the proprietary element of personal data is one of them, which we have done here via the introduction of the proprietary chain of data. The other aspect that needs clarification in order to contextualize personal data in the marketplace is the value of personal data. These are the two necessary elements in a transaction: ownership must be established before it can be transferred, and a value must be established before its price can be paid. Having established ownership here, we must now establish how personal data is valued, and thus how a price is established.

\subsection{The Value Problem: Rivalry \& Excludability}
As part of their excellent analysis of  the challenges and obstacles that personal data markets face, Spiekermann et al crystalize the issue of finding a monetary value for personal data. They argue that personal data has the earmarks of a free commons since it is ``non-rival, cheap to produce, cheap to copy, and cheap to transmit\cite{challenges:spiekermann}". The authors claim that for personal data to resemble a marketplace commodity, it would have to somehow gain the qualities of rivalry and excludability. For personal data to have rival and excludable qualities in the marketplace would require that personal data be placed in a context where having its consumption by one party would mean that no other party would be able to simultaneously consume it, and that only a paying party would be able to benefit from the transaction. Recalling the introductory discussion provided here on knowledge assets, we come back to the inescapable fact that the nature itself of personal data is the primary source of challenges to the data marketplace. Personal data can also be conceptualized as a form of intellectual capital, described by Jean-Jacques Lambin as ``all non-material resources that could be considered as capitalizable assets of an economic agent\cite{Rethinking:lambin}". This means that in order to fully conceptualize a functional data marketplace, it may be necessary to let go of the idea of personal data as a commodity, a raw material that can be sold and traded, and consider its value as an intangible asset.

Lambin is very clear about the pitfalls of intellectual capital, given that they are derived ``directly from the employment of the relational, effective and cerebral faculties of human beings\cite{Rethinking:lambin}". Lambin identifies knowledge as one of the primary members in the category of intellectual capital, and referencing Kenneth Arrow's framework for knowledge as an economic good, also identifies the non-rival and non-exclusive qualities of knowledge as the key components of the ``knowledge dilemma\cite{Rethinking:lambin}". The dilemma highlights the conflictual dynamic between the non-rival and non-exclusive qualities of knowledge, and the great commercial potential of knowledge. For the purpose of this cognitive model when we talk about rivalry we refer to the quality of an asset that keeps it from being consumed by more than one entity. Knowledge assets, as intangible assets, are generally considered non-rival since their consumption by one entity does not keep other entities from also consuming the asset. More specifically, personal data is non-rival since the data generator can sell the same data to more than one customer, the sale does not consume the data itself, although it may have some effect on its value in terms of exclusivity use, which raises the question of the exclusivity of personal data. 

When we talk about excludability we refer to the quality of an asset that does not allow anyone but those who pay for the asset to benefit from it. Generally, knowledge assets are non-excludable because multiple people can benefit from the transaction without being directly involved in it. The data marketplace faces this same dilemma given that its currency is a form of knowledge, knowledge asset, and a form of intellectual capital. In order to answer this dilemma, and in accordance with Lambin's directives, a data marketplace model must find a way to emulate excludability and rivalry to be successful. Without these two qualities, it is impossible to properly determine the value of personal data. A marketplace cannot function if the value of what is for sale is unclear. As stated by the knowledge dilemma, personal data in its raw form cannot readily satisfy the requirements of rivalry and excludability. A personal data marketplace will likely face market failure if it cannot provide a solution to this dilemma, therefore in order to describe how transactions take place, the first order of business for a data marketplace model is to propose a market structure which allows personal data to be rivalrous and excludable. 

Providing a structure that allows personal data to be rivalrous is the actual  crux of the dilemma's challenge to the data marketplace since excludability is met by the virtue of the marketplace transaction itself. Based on a buyer-seller marketplace model, as suggested here, personal data transacted on the data marketplace will only benefit those who pay for it. It is entirely possible that additional parties may benefit from the initial transaction, but that does not remove the marketplace's ability to meet the excludability requirement. These benefits are not proportional to those the data buyer stands to gain from the transaction, they are residual benefits. For example, a possible scenario is one in which personal data purchased on the marketplace is processed by a corporate party and utilized to formulate predictions about customer preferences with the aim of planning a marketing strategy for a new product. A third party, an unassuming consumer completely separate from the personal data transaction, may benefit by having marketing tailored to their tastes. This benefit may manifest by bringing new and interesting products into their sphere of awareness. This type of benefit is a byproduct of the personal data transaction, it does not equate to non-paying parties benefiting from the original transaction. Excludability becomes a problem once again when we consider the post-transition life of personal data. Because the transaction itself does not consume the data, duplication and dissemination is entirely possible post-transaction. This becomes problematic for ensuring excludability is consistent in the marketplace model, and also goes hand in hand with the rivalry problem. Therefore, a deep dive into the nature of the rivalry problem, and how to solve it, is required to move forward with our analysis.

\subsection{Solutions to the Knowledge Dilemma}
What haunts data-based transactions is the fact that it can be replicated at no cost before and after it is sold. The fact that data is not consumed in the process of it being transacted poses a fundamental problem to the marketplace. In order for a data marketplace to meet the rivalry requirement it must then address this challenge. At a high level consideration of this challenge, personal data can be made to emulate a rivalrous good in one of two ways: by artificially forcing its consumption as a step in the transaction process, or by framing it with a structure that places restrictions on its use that simulate rivalry. The first solution can be referred to as the physical solution, since its approach is to physically manipulate the good in question, personal data, in order to change its problematic qualities. The second solution can be referred to as a framing solution, since its approach does not necessarily modify the good in question, but instead it stipulates how the good can be accessed and exploited. We consider the second solution to be the best method for ensuring the data marketplace meets the rivalry and excludability requirements and therefore recommend its incorporation into the marketplace model as a solution to the knowledge dilemma.
	
Indeed, physical solutions, in terms of technological implementations such as Digital Rights Management (DRM), rely on the viability of specific kinds of technology. If such technologies fail, which inevitably happens in a fast-paced technological context, an additional safety net is necessary to ensure the physical solution through a framing solution, like a legal framework. As a matter of fact, the physical solution may only be used as a complement to the framing solution. On the other hand, legal solutions are a strong ground to anchor our marketplace mode, since we do not expect legal systems to become irrelevant any time soon.
	
The framing solution can be implemented by packaging personal data in a licensing framework that includes end-user license agreements to define the rights of each party after the data transaction is complete. Although it may not seem evident, this solution is a natural extension of the user-centered approach that steers the direction of the model proposed here. Licenses serve as both a solution to the knowledge dilemma, as well as safeguards that protect the data generator from unintended uses of their data post-transaction. In order for the framing solution to be effective, its licensing framework must be anchored in clear legal reasoning. The power of this solution withers if the licenses it proposes are not substantiated in any way. We must look at the legal literature to provide the framing solution with a steady anchoring.

There are two schools of thought on the relationship between privacy and copyright: Phillip Hacker's analysis of the legal traditions that inform Intellectual Property (IP) rights in contrast to privacy rights highlights the differences between the European tradition and the American one\cite{ip:hacker}. According to Hacker, in Germany for instance, the legal tradition underwent a ``co-evolution\cite{ip:hacker}" stage during which copyright and privacy laws stemmed from the same philosophical foundations, while this type of coevolution is absent in the development of the US development of IP law. This separation from traditions consummated in a European tradition where IP and data protection became anchored in personal rights, while the American tradition considers IP and data protection to be largely informed by their economic utility\cite{ip:hacker}. Nevertheless, Hacker points out that beyond this theoretical difference there is a rising consensus across both traditions, and the contexts that surrounds them, that understands copyright as grounded in the person, and the inalienable moral rights that protect creative works as they relate to the personality and identity of the person herself\cite{ip:hacker}.

Hacker concedes that the tension between the personality-grounding and the utility-grounding camps of IP discourse does not dissipate entirely based on this newfound understanding\cite{ip:hacker}. His argument is that the rising dissatisfaction with the handling of personal data, on either side of the Atlantic, reveals some truth about the weight of economic utility versus privacy in personal data as perceived by users: that they are at least equal in magnitude. This is not a claim to take lightly, especially considering that those who hold this perception, users who generate personal data, are in fact those very same individuals who are the recipients of the rights that are up for discussion in the first place\cite{ip:hacker}. To sum it up, in Hacker's own words, IP and privacy share a common ground and are ``both economically instrumental and an end in itself\cite{ip:hacker}". The inevitable conclusion that ensues from this view is that marketplaces where IP rights are the subject of transactions can be compared to, and inform the structure of, marketplaces where privacy rights are the subject of transactions\cite{ip:hacker}. 

The model to look at for comparison then is that of an IP regime that is personality-based. This model looks at IP law with the perspective that ``creators" retain a set of moral rights which inextricably link them to their work. Hacker gives the example of the moral rights of authors as they are informed by a continental European approach to copyright law, which includes rights such as the right to publication and attribution\cite{ip:hacker}. This link between an author and the products generated by her creative actions establishes a foundation for a type of property right that informs how ``ownership" of said products is understood\cite{ip:hacker}.  Based on this understanding of IP, we can look at this model for parallel development of a legal regime that provides an understanding of the types of rights that generators of personal data are recipients to. More specifically, we are looking for rights that allow us to establish grounds for a framing solution based on contractual agreements which define licensed access to personal data in exchange for compensation. Thus we draw a parallel between the artists and their creative work, and the data generators and their personal data. Although not evident, there are parallels to draw on for such a comparison.  An author's creative work has a fragment of her personhood inculcated in it, which makes said work unique and identifiable. A dancer's performance, although choreographed, will be different every time, and will be experienced differently every time by a witnessing audience. 

This phenomenon manifests in different forms. A familiar one that demonstrates this effect of inculcated personhood in creative performance is the guitar solo in musical performance: a centerfold to many musical styles and traditions which relies in part on an improvisational performance, unique to every musician. One of the best examples of this phenomenon is the use of the guitar solo in musician Django Reinhardt's unique Jazz legacy. Reinhardt is considered by some to be the ``first true jazz guitar hero\cite{django:mccarty}" based on his incredible skills as an improvisational guitarist. Some have argued that Django's ability as a guitarist was in part a result of an accident that subjected his left hand to burn injuries that both limited and amplified his skill as an improvisational guitarist\cite{django:mccarty}. This anecdote from Reinheardt's life reflects a reality shared by all artists: the uniqueness of the artist's work is inextricably linked to their personal qualities, physical form, and creative faculties. It is tied to the artist's identity, making the work identifiable. This characteristic of uniqueness is in part what can make personal data so valuable. Personal data is similar in the way that it is the result of the combination of things that makes us who we are: be it data generated by our interests, our lifestyles, our occupations, or any other component of our identities. If personal data was not the product of such elements, it would not be identifiable in the first place.  

An inevitable question then arises: how can we compare the mundane activities that often are the spawning point of IoT generated personal data to complex creative and artistic endeavours like improvisational musical performance? The question of ``what is art" has been a subject of debate for a long time. Preconceived notions of what can be called art, and on what merit we can do so, have been refuted by  artists such as Marcel Duchamp, who pioneered the concept of ``readymade" art. Duchamp's view was that the mundane can be elevated to the status of art on the basis that artists themselves give it that title, recognizing it as the result of their creative work\cite{moma:ducahmp}. This dynamic is reflected in the creation of personal data:  it need not be beautiful, unique, or complex to be considered personal. We see a parallel between creative works of art and personal data based on the binding act that generates said data, much like the way an artist claims their work to be art.
 
Moving back from the abstract to the concrete, given the parallels we've highlighted between artists and data generators, what legal concepts can we also analyse in parallel to inform a personal data licensing framework? Hacker suggests that a solution that frames property contractually is the best approach for normatively anchoring personal data in a way that allows it to be subject to the same kind of market transactions that exist for IP\cite{ip:hacker}. The reasoning behind the selection of a contractual solution is the flexibility that contracts afford: they can be shaped to satisfy different requirements in different contexts and can be modified in accordance with the interests of the transacting parties\cite{ip:hacker}.  As Hacker himself puts it: ``a contractual regime implicitly creates a variety of different ways of using personal data, or of excluding third parties from its use\cite{ip:hacker}". This understanding completes our picture of data ownership: the conceptual component provided by the proximity based taxonomy that establishes property entitlements based on the data's instantiation, while the normative component is established via contractual agreements between data buyer and data seller that outline the rights recognised and conditions to which the transaction is subject to, namely permissions to use the data and compensation for said use.

\subsection{The Personal Data License}
The logical extension of this contractual solution in a personal data marketplace is to formalize such contracts as part of the transaction process. Our proposed vehicle for doing so is the personal data license: an end-user agreement that outlines the characteristics necessary to simulate rivalry and excludability. The contract formalizes the willing exchange of access and use of personal data for compensation. The contractual solution then can establish use clauses that limit the access of the data subject to the contract to a single buyer: for instance, the data subject may agree to not sell additional copies of their data. At the same time the contract can establish clauses that limit the use of personal data by the buyer: it may restrict the buyer to specific approved uses of the data, and forbid others. The contract can also establish a lifespan and scope for the access and use of personal data: a defined period of time after which the buyer loses the right to utilize and access the data and the data subject may place said data back on the market. Essentially, the personal data license allows for a normative framework through which the use and access of data can be regulated in such a way so as to replicate the effects of rivalry and excludability without having to physically modify the data itself through technical solutions. This normative framework safeguards both the buyer and the seller against violations of the contractual agreement. In light of this solution, a complete description of the licensing framework, and its primary uses,  follows.

In addition to providing a solution for the rivalry dilemma, the licensing framework can also define and formalize some of the more abstract characteristics of personal data. For example, personal data in its raw form has a number of qualities that can become problematic in a marketplace context. Personal data has dynamic qualities that make  the data valuable only in relation to data from different points in time. For example data about a subject's weight loss over a period of time has a certain value if taken into account within a given context: it's not exactly valuable  to know that at a specific data point on a given day the data subject has a specific body weight. On the other hand, it could be very valuable to know that after that given data point, the subject's body weight decreased. This is especially valuable if the information is  accompanied by data about exercise and diet at the given data points. A licensing framework can capture the value of this dynamic characteristic of personal data and formalize it by establishing an agreement ruling over a spectrum of data points, over time, in the license itself.

At the same time, personal data has static qualities: these are aspects of data that make individual data points valuable in and of themselves based on the possibly unique configuration of data at a given point in time. An example is a data point including a marathon runner's specific blood sugar level the moment a race is completed, alongside their individual bio information such as age, weight, height, etc. This is all static information, its value is crystalized at a specific point in time. Having a large sample of data about blood sugar levels upon completing a long distance race, over a varied population, could be valuable in developing recovery products for long distance runners for example. Some static characteristics can be problematic given that they do not come about via the data subject's activities, and are simply static data generated by environmental and situational forces. Establishing ownership of such static data might be problematic, as static qualities may also give the data shared qualities: meaning that there is information in those static data points that includes information about others besides the primary data subject. The primary example of this is information about the data subject's genetic profile. A data subject's genetic profile necessarily includes some information about the genetic profiles of her parents and siblings. The license framework can establish boundaries on how personal data with shared qualities should and should not be accessed or processed by establishing what kind of activities would constitute non-compliance with the license.

In effect, the data license would formalize agreements based on two main activities performed by the party purchasing the license: access and use. Access details would include information about the regulation of access to the data, essentially the lifespan of the license as well as the level of exclusivity. The primary detail here, especially in consideration of this licensing framework as a response to the rivalry dilemma, is the formalization of exclusive access. This is the great strength of the licensing framework: it can simulate rivalry by granting the license owner exclusive access to the data and precluding the license seller from selling additional licenses to others, while at the same time establishing the resale of the data by the license owner as an act of non-compliance. The advantage of flexibility allows the licensing framework to also grant the license seller the ability to sell additional licenses if their agreement with the license buyer does not require exclusivity to the data. In addition to details about exclusive access, the data license would include details about the lifespan of the access. After the established time period, the license agreements  ruling over access would expire and the license would stipulate what kind of access rights are resumed: the license seller may resume selling licenses and the owner of the expired license may no longer use the data for example. 

The licensing framework's flexibility also allows what in practice equates to ownership transfer of the data. A license that stipulates exclusive access to the holder, while also establishing a lifespan without expiry would simulate direct sale, given that the use of the data as stipulated by the license is appropriate, as well as allow a mechanism to facilitate the right to be forgotten, as stipulated by the European General Data Protection Regulation (EUGDPR) in terms of ceasing data processing once consent is removed\cite{GDPR:EU}. The same way the licensing framework can formalize access rights, it can also formalize use rights. This category of license details essentially describes the scope of the license: what can be done with the data and what can't be done. The scope would prohibit the use of personal data for illegal activities as well as the uses of the data that could put the data subject in danger. The scope of the license would also establish the kind of information that can be extracted from the data. Returning to the discussion on shared qualities, the license scope would establish that information that can be extracted about individuals that are not the license seller themselves is prohibited and non-compliant.

\subsection{Risk as an Indicator of Value}
Up until now, we have given primary consideration to the elements that make personal data valuable for the data consumer: the fact that personal data is a type of knowledge asset that can fuel new economic models. Yet, a conceptual model for a data marketplace would be incomplete were it not to include a significant discussion on what makes personal data valuable to data generators. If this was not the case, data generators would not bother with a marketplace and would be less reluctant to give away their personal data at no cost. There is some discussion about whether it is the case or not that individuals find personal data valuable at all. Psychological models that explain individual perceptions, attitudes, and opinions on the value of personal data are beyond the scope of this paper. The marketplace model we propose operates with the assumption that individuals willing to enter into a transaction with a data buyer recognize that their personal data holds value in an objective way. Where does this value come from? Primarily, we propose that the data generator understands the value of their data in terms of the privacy component that is inherently found in personal data, and the risk associated with the potential violation of  their privacy. In other words, personal data can be considered valuable to data generators because it includes information that can map and expose aspects of the data generator's privacy. Sharing that data is akin to exposure to certain risks by compromising privacy. The relationship between risk and privacy is both extremely complex, and a crucial component of the way personal data is seen as valuable. Therefore, some discussion on this topic is required in order to fully accomplish a complete picture of the value of personal data. 

Privacy is one of the more abstract concepts found in modern society. It is not exactly easy to explain what privacy is, and yet we all understand to some degree of clarity why privacy is important. Furthermore, we all have a strong inclination towards indignation when we recognize that privacy is violated. Often enough it is easier to highlight the importance of privacy through metaphor. For example, consider a locksmith's shop where keys are made. Within the shop, the locksmith keeps a wall where he hangs up keys that have been completed and are waiting to be picked up by a client. Is it a violation of the client's privacy to have the keys exposed to the public in this way? Anyone could walk in, and look at any given key. More than that, they could photograph its shape, or even flat out steal it. Most would agree this is not a problem. Even if a malicious actor were to take a client's key, or replicate it in some way, the keys are not hung up according to any order or pattern. The stolen key, to that malicious actor, holds zero value beyond the value of the metal that composes it. Alternatively, consider that the locksmith hangs the keys up, and labels them according to the client's first name to facilitate delivery. Now imagine they are labeled by first and last name. Perhaps they are also labeled by address. Now imagine they are labeled according to full name, address, and household income. 

It quickly becomes evident that it is not the keys themselves that are the problem, but the access they symbolize. Who wouldn't hesitate to utilize the services of the locksmith with the latter organizing strategy for their completed keys? There is an intangible element which connects the keys to the clients, which we refer to as privacy. Once this connection is intercepted, the clients are at risk of various harms. When it comes to assessing the commercial value of this connection, we traditionally think of risk coverage in the form of insurance. An individual may pay to become protected against the potential damage represented by a risk. We're looking at this relationship in the inverse direction: an individual should be compensated for exposing themselves to potential damage represented by a risk. 

Stepping away from metaphor, Daniel J. Solove provides an excellent description of the benefits of privacy when he claims that ``privacy cannot be understood independently from society [...] privacy is the relief from a range of kinds of social friction [...] it is a protection from a cluster of related activities that impinge upon people in related ways\cite{taxonomy:solove}". This protection is critical to understanding how personal data holds intrinsic value, separate from the value that is determined by the demand for such data from data buyers. When a data generator decides to put their personal data on the market, they forgo this protection and take on risks to a range of harms. These harms are varied, but can be categorized to classify the different risks that data generators take when they enter the data marketplace. Solove has developed a taxonomy that sheds light on a number of ``privacy harms\cite{taxonomy:solove}" that are glaring enough to be distinguishable. 

Originally, Solove proposed his taxonomy to provide legal systems with a bolstered understanding of privacy\cite{taxonomy:solove}. Yet, the broader goal of Solove's taxonomy is to exemplify the distinct problems that individuals face when their privacy is at risk\cite{taxonomy:solove}, which is in line with the goal of clarifying the relationship between risk and value for data generators. We can therefore utilize Solove's taxonomy to categorize privacy risks associated with personal data, and qualify how the risk magnitude can increase based on data types. Solove's target audience does have an impact on the granularity of his taxonomy: courts and legal scholars require in-depth articulation of how to differentiate across many types of privacy harms. We don't need to use the full depth of Solove's taxonomy to categorize risks for our analysis. The logic behind Solove's taxonomy can be used as a reliable backbone to support an attempt at categorizing the risks data generator's are exposed to in the data marketplace. Moreover, the intention behind this assessment is to further highlight the relationship between risk and value: the higher the risk magnitude associated with given personal data, the higher the value it holds for the data generator. 

Solove's taxonomy revolves around the ``data subject\cite{taxonomy:solove}", meaning that the way different privacy harms are grouped is determined by how they affect the data subject\cite{taxonomy:solove}. Solove's activity groupings correspond to information collection, information processing, information dissemination, and invasion\cite{taxonomy:solove}. In the taxonomy, each of these groups is further broken down into ``forms" in which the activity can manifest. For example, information collection can take the form of surveillance activities and interrogation activities\cite{taxonomy:solove}. Leaning on metaphor once more, Solove makes the argument that the activities that the taxonomy groups break down into are separate and distinct, but still very much related, similar to how members of a family share resemblances and traits, yet remain distinct\cite{taxonomy:solove}. What is most salient to our cognitive model from Solove's taxonomy is the undercurrent, or the familiar DNA, that runs under these different activities and connects them. This unifying stream refers to two categorical effects that different activities have on the data subject: dignitary harm  and structural harm\cite{taxonomy:solove}. Dignitary harm refers to those privacy violations that result in damage to the data subject's reputation, or result in damage to the data subject's social environment by adding stress\cite{taxonomy:solove}. Separately from dignitary harms, Solove identifies structural harms as those which do not directly deal with the data subject's social dimension, but instead deal directly with effects on risk of future harm. Structural harm either enhances the likelihood that the data subject will be harmed in the future, or it shifts the balance of power between the data subject and other entities in an unfavourable way\cite{taxonomy:solove}.  

We can further exemplify how the two differ by analyzing a privacy violating activity and fleshing out the different strands of harm in it. Under Solove's category of information processing, secondary use of data is flagged as a privacy harm and is defined as: ``the use of data for purposes unrelated to the purposes for which the data was initially collected without the data subject's consent\cite{taxonomy:solove}". Secondary use poses the structural harm of uncertainty and anxiety, the misuse of this information places the data subject in a position where they have no knowledge about how information about them is being used, and how such secondary uses will impact their lives, leaving them powerless\cite{taxonomy:solove}. In terms of dignitary harms, secondary use of data can result in unwanted identification and distortion of the data subject's identity or reputation\cite{taxonomy:solove}. Secondary use of data can then: reveal information about the data subject that was never meant to be exposed, it can distort the data subject's perception of reality, as well as distort the data subject's identity, and finally it can allow direct intrusions into the data subject's life that were never intended\cite{taxonomy:solove}. Throughout Solove's taxonomy we see a similar pattern, these privacy harms manifest in the same way: they expose the data subject to the risk of revelation, distortion, and intrusion. From this finding, we propose a methodology for categorizing the value of data, for the data subject, based on the magnitude of risk that it poses, based on the ideas behind Solove's taxonomy.

\begin{figure}[t!]
    \centering
    \includegraphics[width=3.5in]{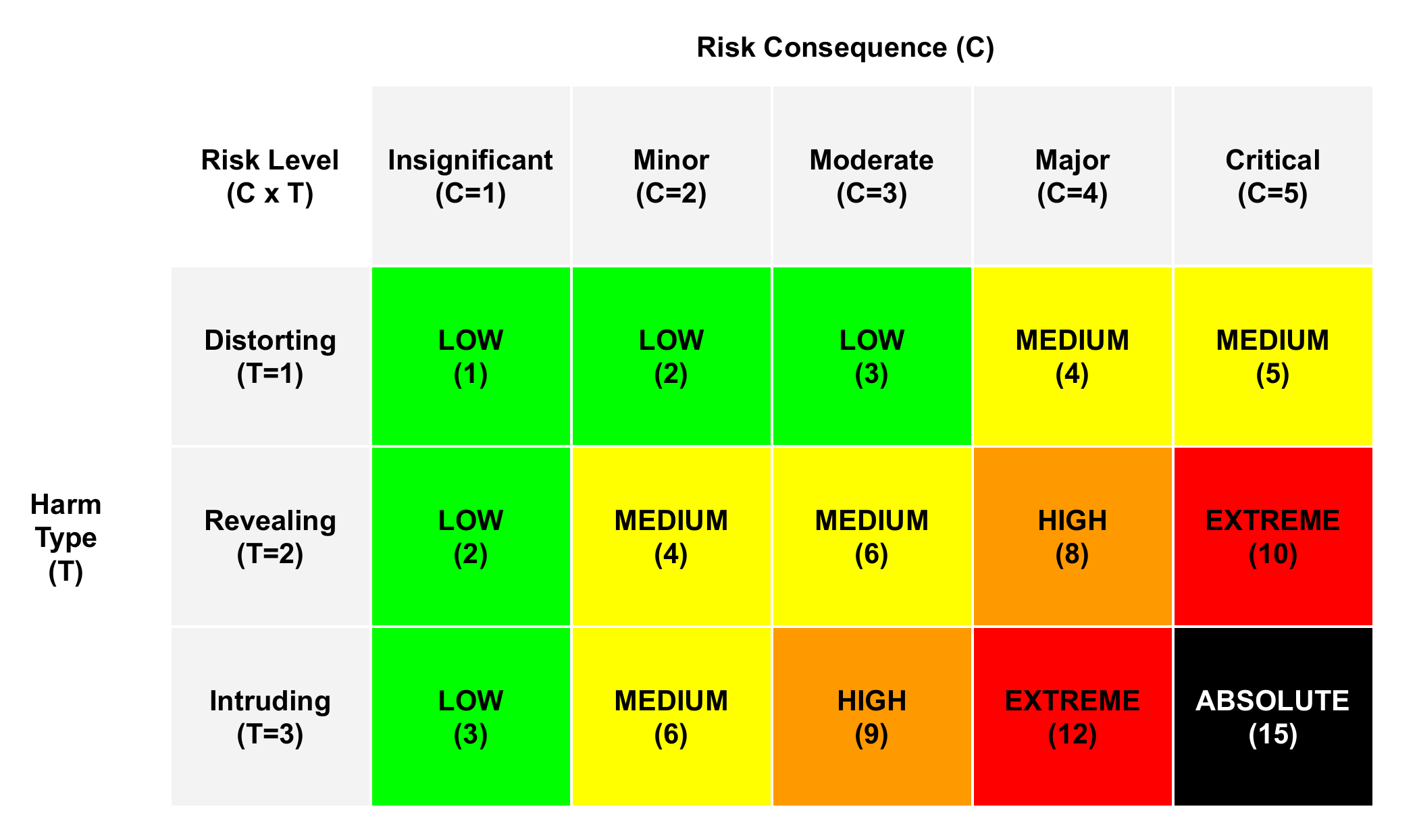}
    \caption{Risk Matrix}
    \label{fig:risk}
\end{figure}
	
\begin{figure}[h!]
    \centering
    \includegraphics[width=2.5in]{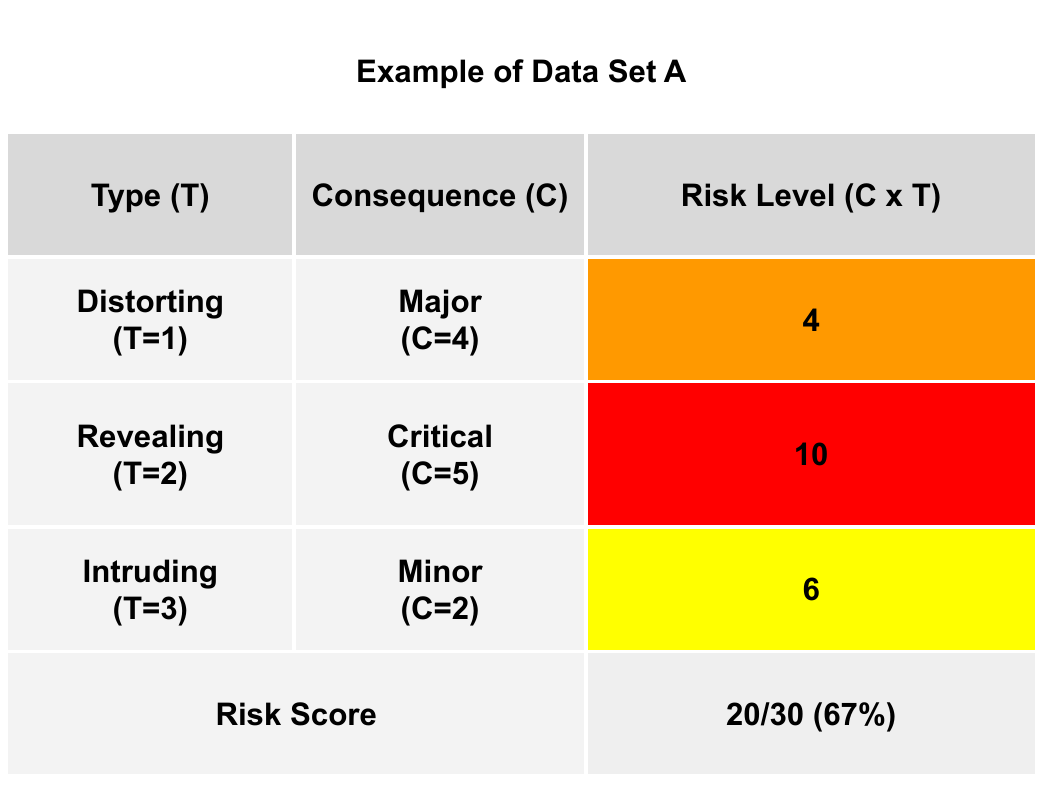}
    \caption{Risk Score Example}
    \label{fig:sample}
\end{figure}

The methodology essentially serves as a mechanism to provide a specific set of data with a tag that associates that data set with a level of risk based on where it aligns according to a risk matrix. This risk matrix is derived from the previous discussion on harm types and their correlated risk impact (Figure~\ref{fig:risk}). The matrix is formed by intersecting cause and consequence: harm type as a cause on the matrix's vertical axis, and impact of risk as a consequence on the matrix's horizontal axis. Harm type is layered. Distorting impact, revealing impact, and intruding impact are each an aspect of impact that a given privacy exposure can have on the data subject. Harm type ranges on the matrix from a sequence of harm escalations beginning with distortion (T=1), followed by revelation (T=2), and finally intrusion (T=3) as the most severe harm type. Type acts as a multiplier based on severity, intruding effects are three times as severe as distorting, revealing are two times as severe, and distorting are not multiplied. Impact of risk ranges according to a spectrum: from low impact (C=1) to critical (C=5) impact. The matrix serves as a tool to categorize personal data sets by the risks that they exposes the data-subject to, ranging on a scale from. The end result of applying the matrix to a data set is a ``risk tag" that associates a data set with a  risk score (R) calculated by adding the individual risk levels at each type layer and dividing it by the maximum score of 30, the maximum being where each type has a critical consequence. For example, Data Set A (Figure~\ref{fig:sample}) may expose the data-subject to a major risk consequence of distortion , a critical risk impact of revelation, and a minor risk impact of intrusion. This data set would then be assigned a risk score of 20 out of a potential 30, or approximately 67\% risk. A data set's risk score can then be used to assign a risk value to the data set, which ultimately contributes to the valuation of the data set itself based on the magnitude of the risk it symbolizes for the data subject.

\subsection{Risk Modifiers}
Having a strong understanding of risk magnitude using the risk assessment methodology outlined in the previous section, we turn to consider mechanisms that a personal data marketplace can introduce in order to mitigate high risk. These mechanisms take the shape of risk modifiers that lessen the effects of high magnitude risks in data transactions. The core strategy inside the risk modifier mechanism is to implement risk mitigation techniques to modify the risk associated with some data, therefore also affecting its value. This mechanism comes into play when we consider specific instances of high risk and high potential harm. Essentially, it is most relevant in situations when considering the threat of malicious marketplace actors that seek to use the inherent risks built in to personal data transactions to harm or exploit data subjects in any way. 

One of the primary concerns regarding privacy within the context of selling IoT generated personal data involves the idea of intended and unintended use of data, specifically with the involvement of an adversary: a theoretical entity with a malignant intent seeking to utilize purchased personal data to draw dangerous correlations about the user that generated said data. Because we are considering the sale of personal data in a marketplace, the angle that we will take on privacy here is not concerned with the danger of a possible data breach and its repercussions. The threat of a data breach and the danger it poses on privacy is common across all systems that manage such data, therefore it is crucial that we drill down to the privacy threats that are specific to a model in which users put their own data up for sale willingly. As we have already mentioned, this is a model that assumes that personal data will be purchased and put to use. The model should therefore safeguard users from a potential adversary who seeks to purchase data with the intention of compromising the user's privacy for gain. We have already discussed intent when we explored data licenses, and the different permissions for data use should be explicitly stated in the data license agreement. Nevertheless, an additional mechanism must exist to safeguard against malicious actors if we seek to protect the privacy of data generators in the marketplace. The personal data license is not meant to be a deterrent against malicious activity, its purpose is to formalize transaction agreements and protect personal data rights. Indeed, the data license can operate as a risk modifier via the effects of the use and access clauses of the license. For example, a license that allows data to be re-sold introduces additional risk to the transaction and thereby has an effect on pricing. 

The threat of a malicious actor in the marketplace seeking to purchase data with the goal of exploiting the privacy of sellers on the marketplace is a distinct scenario, which requires its own discussion. We suggest a two-pronged strategy for this purpose: one approach is active and the other is passive. The passive approach is a reputation system for the data marketplace. This kind of mechanism actually serves a dual purpose: it protects buyers and sellers from malicious actors equally. It is not a mechanism specific to the scenario with malicious data buyers seeking to exploit the privacy of data generators. We will discuss this reputation mechanism at length in the description of the marketplace model in the next section of the paper. On the other hand, the active approach we suggest here is directly relevant to the malicious buyer scenario. 

We have discussed at length how personal data is a challenging and complex knowledge asset, lying at the intersection of many domains. Personal data, categorically, exists in an overlap between private information, thus exclusive to a certain extent, and commercially useful information, therefore desirable. This categorical overlap gives personal data the potential to transform industry and fuel innovation for those that can utilize it, but it also gives it the potential to devastate the privacy of those that generate it. We must differentiate the two, and somehow protect privacy without hindering utility. This is difficult because they both operate in an overlap, therefore modifying one has an effect on the other. For example, a user sells a batch of data generated by walking across a variety of terrains to a company that design shoes. The company purchases the data with the aim of using it to optimize their sneaker design. Our marketplace model must protect certain aspects of the user's data. In this case, the data should reveal the details necessary for the optimization of the company's sneaker design: it should allow the company to determine the effect of an incline on the stamina of the user generating data walking a given distance, but it should not allow the company to determine the address or work location of the user generating the data via GPS coordinates.

Risk modifiers deal with both categories, they modify the data to protect privacy, and therefore lessens risk. At the same time, they also diminish the value of the data by decreasing its utility. We need to consider the practical implications of such mechanisms in the marketplace as they are implemented to ensure they do not undermine the overall efficiency of the marketplace.  Currently, the convention agreed upon to address this issue is the introduction of noise into the data in a careful manner, so as to retain the useful components of it for the buyer, but without revealing information about the seller that an adversary could utilize\cite{newDataMarkets:dong}. This convention hinges on one thing: it must be determined how much noise can the buyer tolerate before the data is rendered useless for them. 

This ``utility-privacy tradeoff\cite{newDataMarkets:dong}" can become a considerable obstacle to the use of data in a context where we are aiming to keep user privacy intact while maximizing data utility. How do we find a balance within the tradeoff?  A possible answer is offered up by the data marketplace itself: if we treat the degree of noise within data in the market as a variable that can be adjusted, it can be considered a risk marker that is taken into account when determining the monetary value of the data. The users are given the ability to set the ``noise levels" within their data, based on recommendations informed by the risk score generated from the data, therefore protecting their privacy before selling. Buyers are given the ability set their noise tolerance as they search for transaction partners, therefore protecting their ability to utilize data by signaling how much noise they are willing to tolerate before buying. Introducing noise into the data devalues it, effectively lowering its listing price on the marketplace. Noise free data then becomes very expensive, and highly rewarding for sellers at the expense of full privacy exposure. On the other hand, data that has noise introduced into it to protect the seller's privacy is more affordable, therefore more attractive to buyers, and safer for sellers. 

The point where noise tolerance meets a satisfactory price is the balance between utility and privacy as it is set by the marketplace itself. This way, a theoretical adversary would find their ability to draw correlations to be blocked by the noise introduced into the data, and would be forced to pay an unacceptable price to obtain the data without distortion, these ``noise-free" premiums would eat into whatever theoretical profit margin would result from their attempts to manipulate the data marketplace and profit from compromising users' privacy. Essentially, noise as a risk modifier provides a mechanism that allows the marketplace itself to find an equilibrium point for different kinds of data in terms of what is an acceptable transaction price for a given pairing of noise and utility. We can take this kind of equilibrium point as an inspiration to introduce how reputation, as a passive risk modifier, can work in the marketplace. Having a reputation score for buyers and sellers can also be an indicator of risk, transacting with a low reputation partner is riskier, therefore it can also serve as a market force to establish an equilibrium point for pricing. Reputation can operate as a risk modifier because it can protect privacy, lessening risk, and have an effect on pricing. Take an exchange between a buyer with a low reputation buying data with a high level of noise modification. The high level of noise will likely reduce the value of the data, therefore lowering its price, but the low reputation of the buyer will impose a risk premium on their end as well, thereby bringing the actual price to an equilibrium point where the seller is properly compensated while their privacy is protected.

\subsection{Personal Data Value: the Full Picture}
Our original analysis goal, and a requisite step to later describe the dynamics of our marketplace model, was to define what personal data looks like in a marketplace context. We identified two components for this definition: ownership and value. We defined ownership in a marketplace context based on a taxonomy of proprietary proximity. Defining value proved a more difficult task given it required an assessment of the problematic  qualities of personal data: namely its lack of rivalry and excludability, and its inherent risk components. In order to clarify the value of personal data, we decided to embed personal data within a licensing framework, and to tag it through a risk assessment that identifies its risk components. By doing so, we have established two aspects of the definition of personal data in a marketplace context that can inform its value: the license characteristics of a given set of personal data, and its given risk assessment. With these two in mind, we can return to the introductory discussion from the beginning of this paper to complete our picture of personal data value in a marketplace context. 

\begin{figure}[!h]
    \centering
    \includegraphics[width=2.5in]{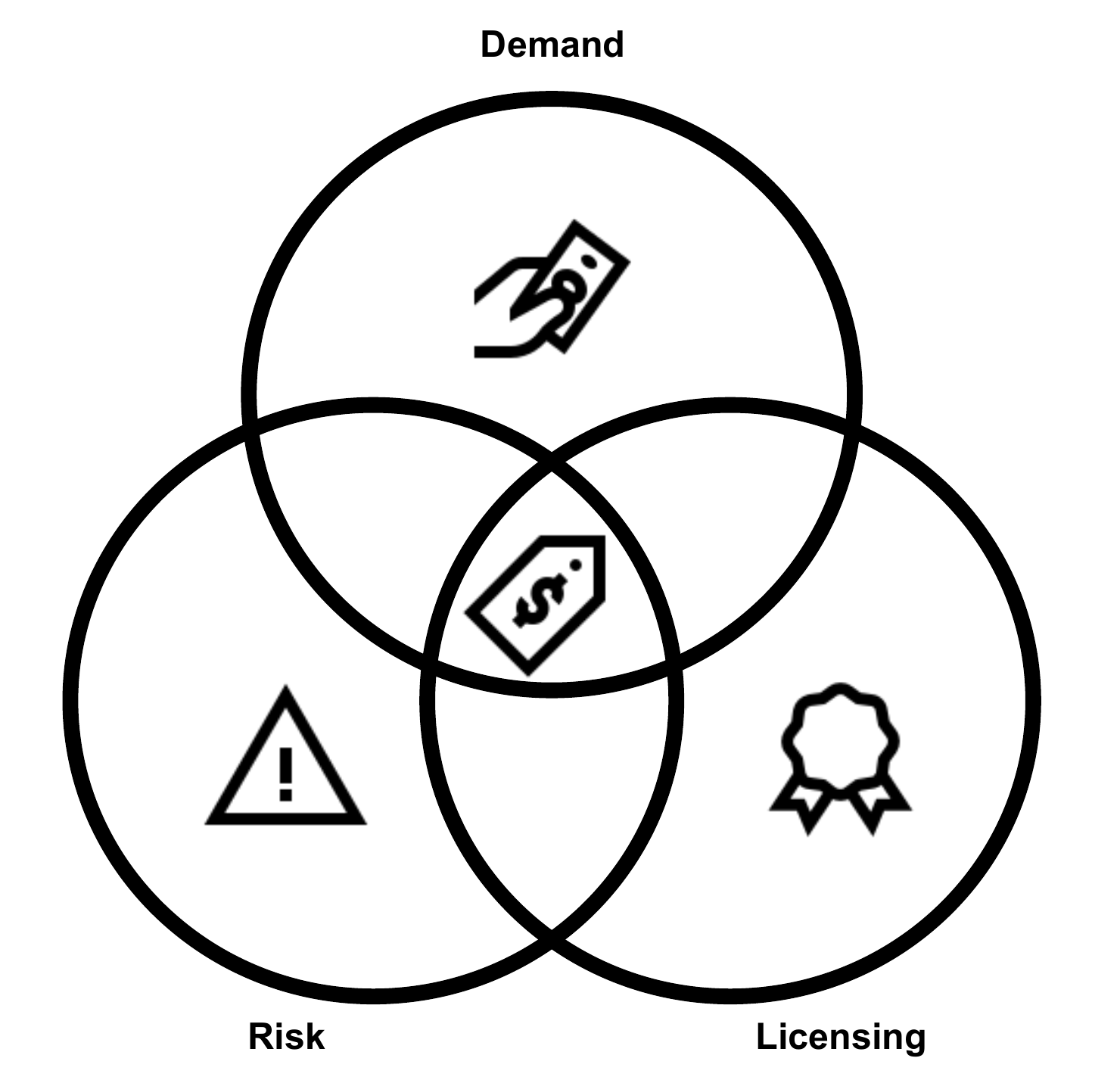}
    \caption{Value Model}
    \label{fig:value}
\end{figure}

We can begin to see the full picture of our value model take shape (Figure~\ref{fig:value}). The effect of licensing and risk on value are components that inform our understanding of data value from the perspective of the data generator, but we must also consider how market demand informs the value of personal data. Modelling market demand is beyond the scope of this paper, we operate with the assumption that there is demand exists based on the implications of industrial advancement using personal data. Even though it is an assumption of our analysis, it would be a mistake to not consider its effect on the value of personal data. Our definition of personal data value is a model that helps clarify where the market-level price for personal data should eventually come from. Making market-level price predictions is beyond the scope of this paper given the variety of personal data categories and circumstances that can be thought up. Instead, our personal data value analysis serves as our model for informing marketplace transaction dynamics with the understanding that a final market-level price for personal data should be determined by licensing, risk, and demand. In this model, licensing corresponds to the specifications of data access and use associated with a given product listing, risk corresponds to the risk value alongside risk modifiers applied in that product listing, and demand corresponds to the actual market demand for a product with the aforementioned specifications.

\section{Proposed Marketplace Model \& Interactions}
\subsection{Goal of Data Marketplace}
Now that we have defined the marketplace context as well as its dynamics, we can move on to a full fledged description of how our proposed marketplace model operates. Previously we focused on the definition of personal data in a marketplace context to give a clear understanding of the requirements of marketplace transactions. This definition yields the data product that is the subject of our marketplace model and its interactions. This section describes the process through which marketplace transactions are resolved, namely the transaction flow, while addressing the challenges that have become relevant as part of our initial analysis. Our proposed marketplace model includes the model components we have described already, as well as introduces  additional components that complete the model.

\subsection{Marketplace Model Description}
Our proposed marketplace model is a digital platform that facilitates personal data transactions. In our previous analysis we outlined a series of model components required to address some of the anticipated obstacles the marketplace would face due to the challenges posed by placing personal data in a marketplace context. Given that the marketplace model includes such model components as risk assessment and modification, and license generation, we recognize that the marketplace model will include some type of agent to monitor tasks required by the model's implementation of said components. This brings us to a defining point in our description of the marketplace model: how heavy handed should this authority be in its oversight of the marketplace model, and what kind of effect does this have on the model itself?

The question comes down to a decision between a centralized or a decentralized marketplace structure. It is true that centralization exists in a spectrum but a critical distinction exists between decentralized and centralized structures regardless of their location on a spectrum.  What we mean by a centralization is a structure that requires activity to transpire exclusively through a given medium, interactions between different stakeholders are all monitored by a single actor or organization. We see decentralization on the other hand as a structure that allows direct interaction between stakeholders with little or no mediation. The differences between these two organizational characteristics has a number of effects on the marketplace model on a practical level. These differences are particularly evident in the diffusion of responsibilities and expectations from marketplace stakeholders.
	
This decision between structures would result in two different kinds of marketplace models. The first kind, a centralized marketplace, would be a data marketplace where a marketplace orchestrator has complete control over transactions: buyers and sellers do not interact with each other. In a centralized marketplace model, data generators submit their data to a centralized repository in the marketplace, and data buyers query the orchestrator with their specific demands for data consumption. The orchestrator then responds with a suitable candidate from its inventory of personal data, collects payment from the buyer and compensates the seller. This is a rough simplification of how the transaction may actually transpire and includes no discussion on the software and hardware architecture required and utilized for a transaction to be resolved. Although such specifications are important, they do not change the essence of how a centralized marketplace transaction would take place, which is what we describe. 

On the other hand, a decentralized marketplace model would be where data generators and data buyers do interact. In a decentralized marketplace buyers can select from listings that are made available by sellers. Once a buyer finds a listing they are interested in, a monitored exchange takes place between stakeholders with the help of the orchestrator. Again, no software or hardware specifications are necessary just yet to clarify how this transaction unfolds.  We see a decentralized marketplace model to be in line with our central mission: making the marketplace user centric and giving data generators the necessary latitude to protect their privacy and data rights. There is a monitoring role to be played by the marketplace, and we have already highlighted what responsibilities that role may entail, but fundamentally the dynamic of the marketplace is one between peers with little mediation by the orchestrator. 

Specific architectures have been proposed elsewhere for data marketplaces. Although we don't propose a specific technology for implementing our model, there is a benefit in discussing examples of what kind of architecture could be used to implement what we propose here. Roman and Stefano provide a reference architecture for a trust-enabled data marketplace that is decentralized\cite{reference:roman}. Much like ourselves, they recognize the disadvantage of current centralized systems. They argue that centralized marketplaces have heavy and often obtuse access processes, including terms and conditions and other agreements, which make participation difficult and take control away data generators\cite{reference:roman}. Their reference architecture is designed to allow different parties to interact in a distributed manner, with some assistance from an orchestrating agent\cite{reference:roman}. Their implementation is based on the use of homomorphic encryption in tandem with Multi Party Computing in an encrypted cloud environment to allow multiple different data consumers to access and process data aggregated and stored by data generators, using a Blockchain to process payments and authorizations\cite{reference:roman}. The architecture suggested by the authors could easily be modified to be used as a reference for the implementation of our proposed model. We have decided to remain quasi-agnostic regarding a specific technological implementation for our proposed model to emphasize its flexibility as a conceptual model that can be adopted under different contexts and towards different ends. Ultimately, we do recognize the large advantage that using distributed architectures and privacy enhancing technologies bring to our model, and include them in our description.

\subsection{Marketplace Actors}
With a clear picture of the kind of marketplace model we have in mind, we can describe the marketplace actors. Our marketplace model involves three actors: data generators with the role of sellers, data consumers with the role of buyers, and an orchestrator with the role of monitor. Each actor has specific motivations, goals, and concerns linked to their participation in the data marketplace.

\begin{figure}[!h]
    \centering
    \includegraphics[width=3.5in]{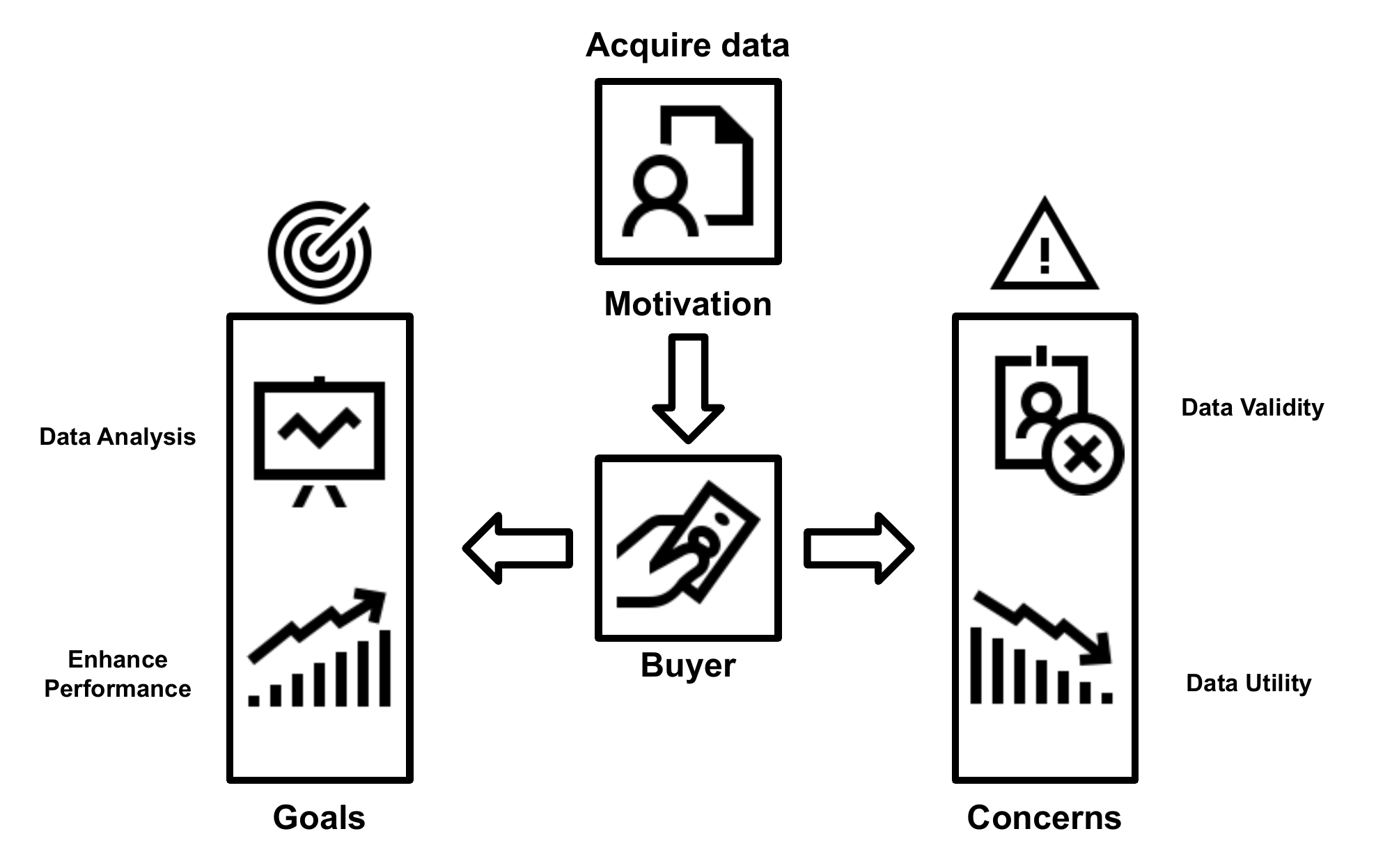}
    \caption{Buyer Model}
    \label{fig:buyer}
\end{figure}

Data buyers (Figure~\ref{fig:buyer})\footnote{Icons by https://icons8.com/, used under CC license} are actors who have the intention of purchasing data. Their primary motivation is to acquire data that can be subjected to data processing or algorithmic analysis to generate insights that can inform their activities. Based on the current commercial climate highlighted earlier in this paper, the current understanding of such an actor is a corporate entity that is interested in gaining commercial advantages through these insights. Their goal may be to use information distilled from personal data to craft new products, tailor their services to a specified customer base, or modify their marketing and outreach strategies, as well as other kinds of activities. A generalized goal exists across these activities that can be applied to the category of the data buyers' diverse goals: to harness the utility of personal data to improve performance. In response to this motivation and goal, data buyers' concerns regarding their participation in the marketplace is the requirement of a guarantee that the data they acquire will have the utility they expect. This guarantee must fulfill two criteria: a demand for validity and a demand for exclusivity. This concern is a manifestation of the adverse selection problem in the sense that it is the result of an information asymmetry between buyer and seller. The marketplace must address this asymmetry to keep data buyers motivated to participate, therefore some form of data validation mechanism is necessary, as well as a mechanism to determine the degree of exclusivity that buyers and sellers are willing to transact over. The role of the data buyer is to specify the kind of data they want to acquire, as well as information about the price they are willing to transact under and their exclusivity requirements according to their utility expectations.

\begin{figure}[!h]
    \centering
    \includegraphics[width=3.5in]{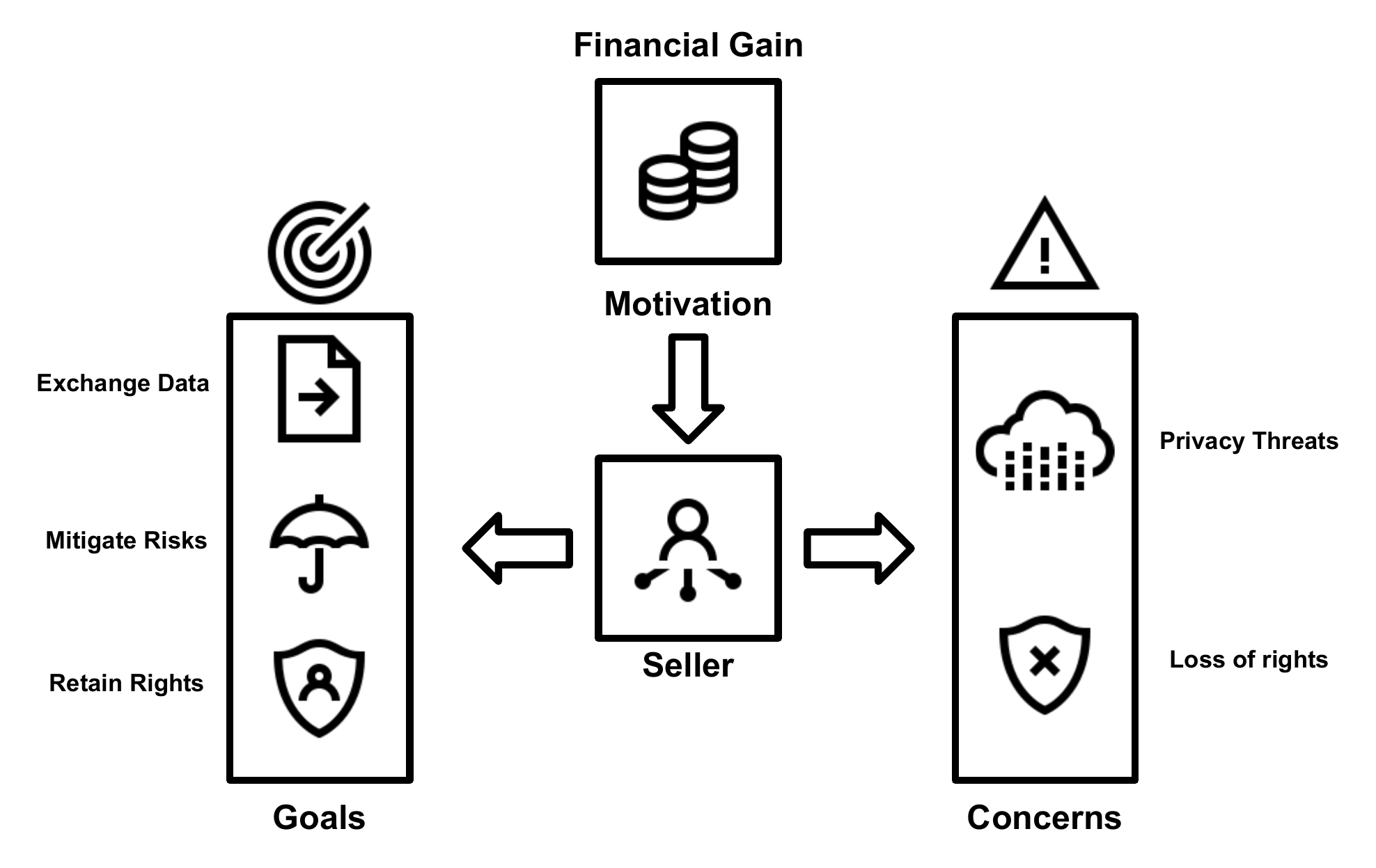}
    \caption{Seller Model}
    \label{fig:seller}
\end{figure}
	
Data generators (Figure~\ref{fig:seller})  are actors who are interested in selling access to their personal data. Their primary motivation for participating in the marketplace is financial gain as compensation for their data. The goal of data generators is to successfully engage in marketplace transactions that result in their financial compensation without sacrificing privacy rights or control over their data, while mitigating the risks associated with personal data transactions. We have already established the risk and value components of personal data earlier in the paper, which are the concepts that inform the data generator's goals. A market value for data is also relevant, as indicated by the willingness to buy that exists via the data buyer's motivation, so the data generator's goals lie on a balance between three weights that determine the value of personal data: risk, licensing, and demand. The goals of data generators compete with the goals of data buyers on the privacy front: their concern is that their privacy and control will diminish in importance under marketplace dynamics. This concern is a manifestation of the expropriation problem: it is a result of the power asymmetry between buyer and seller based on the low cost of data replication. Since there is no cost associated with replicating the data, it is likely that data generators will lose all control over their personal data, and its uses, once it is sold. The marketplace must address this asymmetry as well in order to keep sellers engaged. The role of the data generator is to specify their willingness to transact, including information about the price they are willing to transact over as well as the specification of access and use.

\begin{figure}[!h]
    \centering
    \includegraphics[width=3.5in]{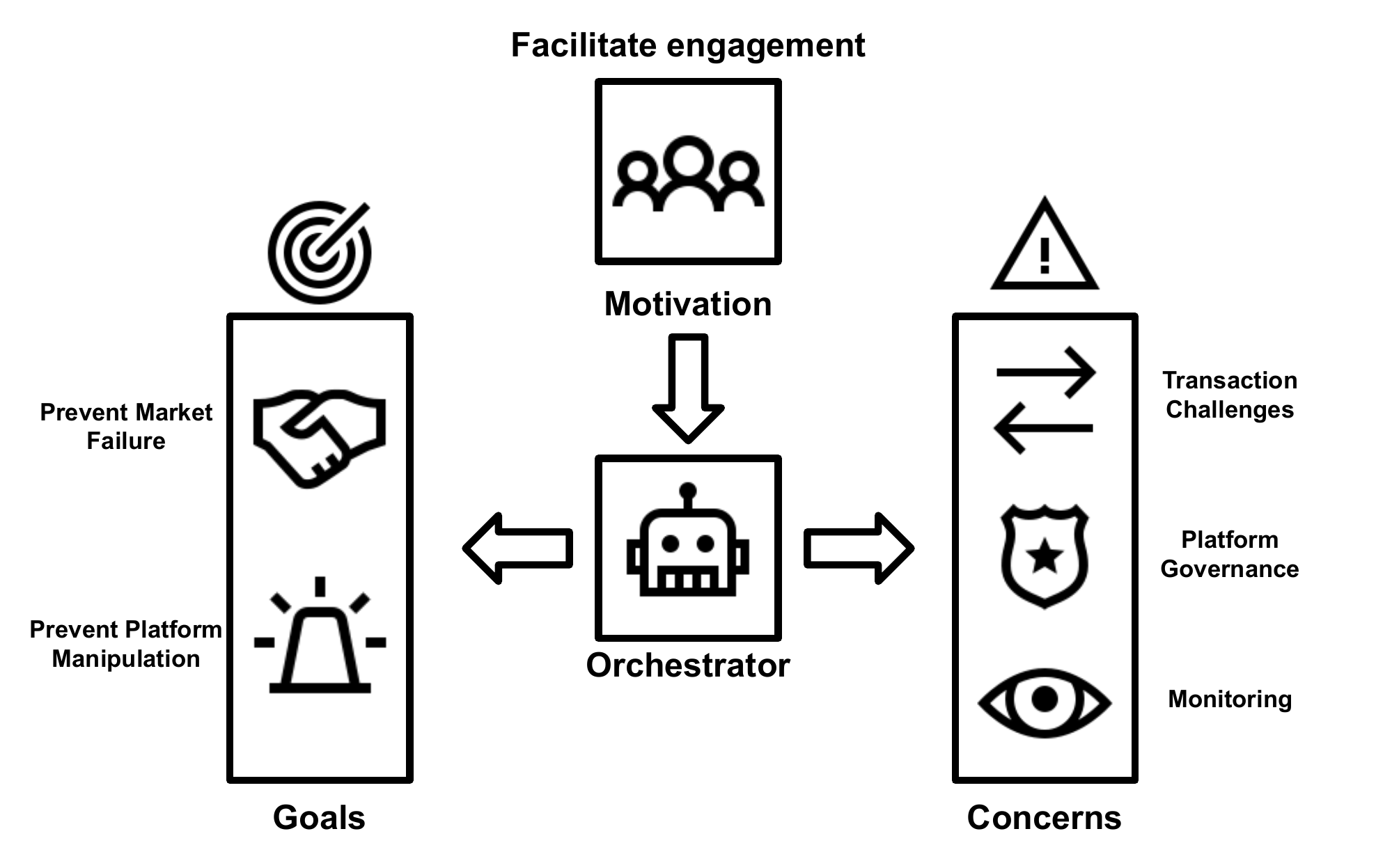}
    \caption{Orchestrator Model}
    \label{fig:orchestrator}
\end{figure}

The last actor to be described is the orchestrator (Figure~\ref{fig:orchestrator}). This kind of actor is different from the data generator and the data buyer. The orchestrator does not need to be an actual person, it can be a set of mechanisms embedded in a system architecture or a software. Even if it is a non-human entity, its design must follow a set of motivations, goals and concerns. The orchestrator has the single motivation of keeping the other two stakeholders engaged in the marketplace. Essentially, the orchestrator has the motivation of resolving the concerns of the two other actors. The goal of the orchestrator is to avoid market failure: a situation in which no seller or buyer can engage in efficient and beneficial transactions due to different obstacles.  The orchestrator's concerns are clearly defined by what hazards we have highlighted that can pose significant threats to a personal data marketplace. 

Four kinds of hazard are evident: challenges rising from asymmetries amongst actors, personal data challenges rising from the nature of the subject of exchange in the marketplace, governance challenges rising from the legal context which personal data transactions take place, and behavioural challenges that are involved in any model where actors may have competing interests. Therefore the orchestrator has the role of resolving or at least mitigating the effects of the adverse selection and expropriation dilemma, as well as the rivalry and excludability problems. In addition to these two marketplace and data challenges, the orchestrator must also monitor behaviour and enforce the policies according to which the model is governed. This role includes facilitating interactions between buyers and sellers to mitigate information asymmetries, maintain the licensing framework that keeps personal data rival and exclusive, enforce policies that would lead to abuse of the model or criminal activity, and finally monitoring against marketplace manipulation.

As mentioned before, the orchestrator does not need to be a person. As a matter of fact, the marketplace model we propose here has a distributed orchestrator: essentially an aggregate of monitoring activities across the transaction process that ensure marketplace policy is followed. The cardinality of the orchestrator is not necessarily limited, at any given point in time different entities can act as an orchestrator as they monitor necessary marketplace functions. Our vision is that of an authority, as a single unit or an aggregate of units or mechanisms, that can act as a digital notary witnessing and validating different functions of the marketplace, and signaling for necessary marketplace actions under specific circumstances. Our proposed implementation for this orchestrator role relies on open standards due to their inherent benefits of balance and openness, as well as interoperability and potential for innovation\cite{openstandards-open}.

\subsection{Valuation \& Pricing}
The value of personal data is the central motivating force behind the actors in the marketplace. The next logical step after having a model that explains how personal data is valued is to determine how value translates to price in order to move forward to a transaction agreement in the transaction process. A successful transaction requires that both parties agree on a price to transact over, therefore we must elaborate on how our value model can produce a price for a given transaction. We identified three determinants of data value in our definition: demand, licensing, and risk. Our value model operates as a pricing function that takes actor behavior as an input, and returns a price as a result. This function is conceptually equivalent to our three-part value model and is only different in terms of speaking of a pricing function as the actual implementation of our value model at a marketplace-level. Describing how the value model works alongside our descriptions of marketplace actors help determine a price requires an overview of how these components function, as well as some new insights into how they interact.

As per the model definition, the value is determined by three elements: information about risks, information about demand, and information about licensing. This information is provided by the different goals, concerns, and motivations that we have highlighted in the actor models. Correspondingly, our pricing function allows us to fully explain how the value of personal data in the marketplace is a result of marketplace behaviors and decisions. These behaviours and decisions are informed by the qualities and asymmetries of personal data that we have taken the time to discuss at length in the first part of this paper. Considering our complete view of our value model, there is a final step that separates value from price, namely details that are not unique to personal data itself as determinants of value. As the pricing function takes input from the behaviour of marketplace actors, and outputs a pricing estimate, an additional consideration must take place to translate this estimate into a market level price recommendation. Transaction details such as quantity of data allow a final price recommendation since the valuation of personal data as done at a single-unit level.

The first input component of the pricing function is demand. In our original description of the value model, we inferred that demand was an assumption made based on contemporary views on the potential of personal data and projections of future uses. Now that we have defined the actors that engage in the marketplace though, we can look at how their behaviours actually would affect demand. The two relevant aspects of the marketplace actors in relation to demand are the motivations of buyers and sellers as described by their respective models. Buyers are motivated to acquire data and in turn sellers are motivated by financial compensation in return for access to their data. The feedback between these two behaviours establishes the demand aspect of the value model, and the way in which they can differ has an effect on pricing via the variety of inputs into the pricing function that these motivations can take. For example, if not enough buyers are motivated to acquire a specific kind of data, or not enough sellers are motivated to seek financial compensation in exchange for a specific kind of their personal data, this will have an impact on the price of said kind of data, following traditional supply and demand marketplace dynamics.
	
The second input component of the pricing function is risk. We established a risk matrix earlier in the paper to determine the risk score of specific kinds of personal data on the marketplace. Now that we have defined the concerns that make this risk matrix relevant for actors, we can see how risk scores can have an effect on the price of data. We know that data buyers have as one of their goals the utilization of data for enhancing performance, which requires valid personal data. At the same time, buyers have the concern that invalid data will not lead them to false findings, and not be an asset in terms of enhancing their performance. In turn, data sellers are concerned that allowing access to the details of their personal data exposes them to privacy harms. The resulting dynamic between these two types of behavior is a balance between privacy and utility, as it is determined by risk modifiers. The risk matrix helps establish the extent of privacy harms in a categorical manner, providing a framework to assess how a tradeoff between utility and privacy can be managed using modifiers themselves. Privacy harms can be mitigated by modifying data, while affecting its utility. Therefore, the impact of risk modifiers on the price of data will be considerable, as we have already established.
 
The final input component of the value function is licensing. Licensing represents a number of key factors in the model. It is a mechanism through which we can address the lack of native rivalry and excludability in personal data as a product. It allows data buyers to access the data they require to meet their goals, and at the same time it allows individuals to retain some say over what happens with their personal data. It is a bridging mechanism necessitated by the tradeoff between access and control. The personal data license establishes an understanding between buyer and seller that stipulates what kind of access and use is allowed, and what kind is not. The details included in this agreement have an impact on the value of the access to the data, and therefore its price. Similar to any kind of license, an agreement that allows a longer period of use will generally cost more than an agreement that limits the amount of time a data buyer has to utilize their access for maximum benefit. Additionally, licensing can have other impacts on the value of the data access based on further limitations. For example, a license that allows the access to be shared with second parties may be more costly given that it requires a data seller to consent to a less restrictive agreement. The behaviours that act as input into the licensing component of our value model come from the remaining goals and concerns of data buyers and sellers, namely the buyer's need for access and use, and the seller's need for control. The variations in these two inputs help inform the price that our pricing function provides for a type of data, embedded in a licensing agreement.

\subsection{Transaction process}
Up until now we have focused on challenges of a personal data marketplace, and have crafted a series of model components required to address these challenges. Having fully delved into the risk assessment, the data license, the pricing function, and the actor models, we can put them together in a practical explanation of how the marketplace operates using these components. This section serves as an explanation of a standard data transaction taking place in the marketplace. The transaction process incorporates three actors: buyer, seller, and orchestrator. It is important to recall that this is a decentralized online platform, therefore there is little mediation in terms of the interactions between buyers and sellers. Our model demonstrates this, where the orchestrator assists buyer and sellers as well as monitors marketplace behaviour as necessary. An additional thing to keep in mind before we begin the description of the transaction process: this description only incorporates one buyer and one seller, while in reality there are multiple buyers and sellers taking the same actions across the marketplace.
 
The transaction process is composed of seven functions: identification, specification, market search, product generation, market match, subsample analysis, and exchange. A full description of each phase follows, but the essential progression of a transaction flows via the resolution of these functions, and the decisions they prompt. The transaction flow as described is sequential, but not necessarily linear. Each of these functions must take place sequentially at some point in time for a transaction to be completed, but it does not mean that every transaction needs to execute linearly through each function every time.The path of buyers and sellers cross in terms of motivations, concerns, and goals, but they require consensus on decisions only twice for a transaction to result in a successful exchange. If a decision results in unwillingness to proceed to the next function in the transaction flow buyers and sellers may reconfigure their specifications as the marketplace dynamics change. Once a buyer reaches a decision point, they may decide the price to be too high. They can return to their specifications to reformulate their needs and search accordingly to their budget, or they can return to a new market search and keep looking for the same kind of listing to appear at a lower price. If a seller finds their listing remains on the market for too long, and no interest is shown, they can return to their specifications and reconfigure them to better suit the market.

\begin{figure}[!t]
    \centering
    \includegraphics[width=3.5in]{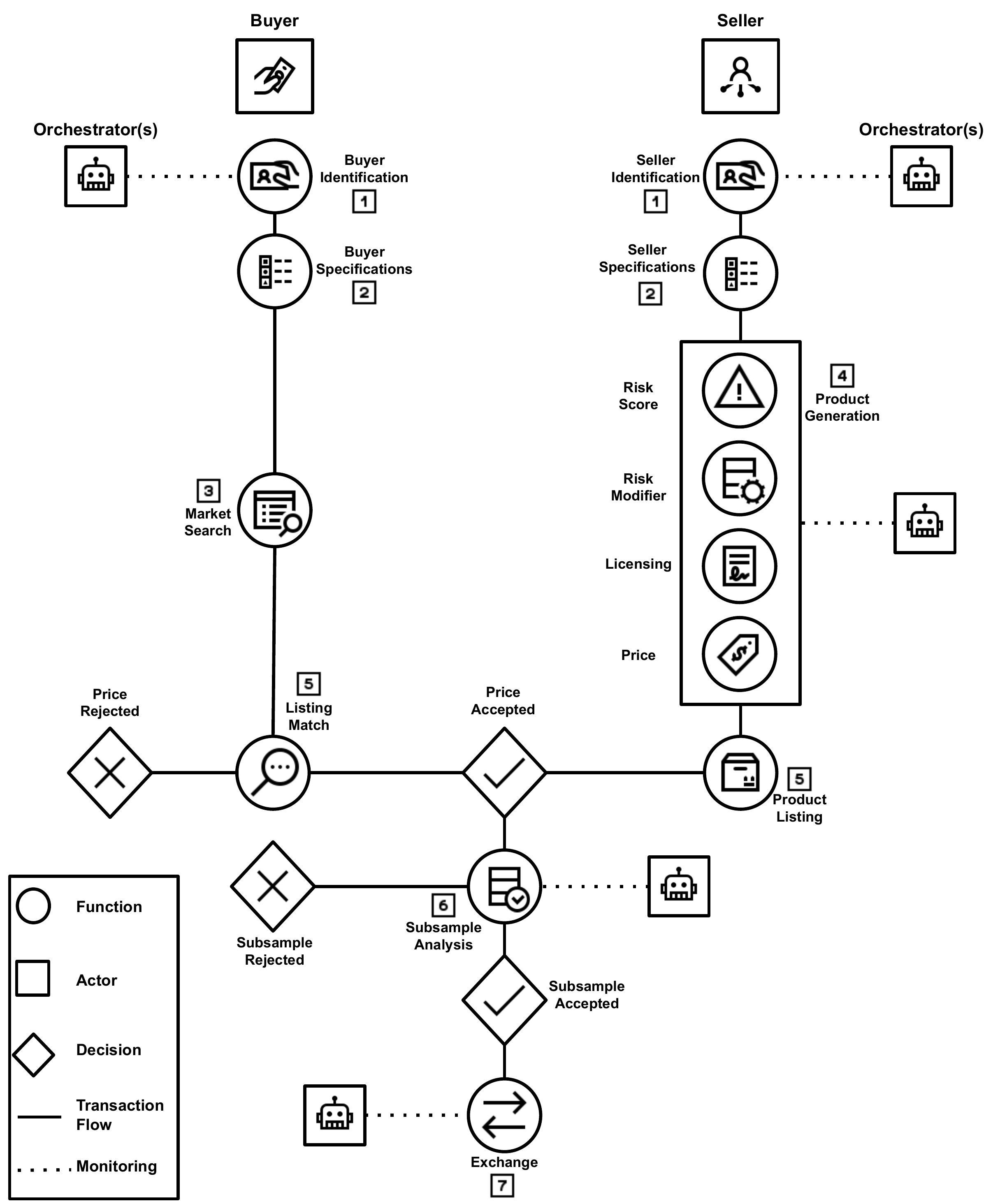}
    \caption{Transaction Model}
    \label{fig:transaction}
\end{figure}

The first function of the transaction process is identification (1 in Figure~\ref{fig:transaction}). This is the point at which buyers and sellers enter the marketplace either as new members, or if they are returning members simply return for a new transaction. The identification function itself is comprised of a number of events. Initially, the buyer and seller have their identity authenticated as members of the marketplace. This is likely implemented through a login system, not unlike most popular digital platforms. Specific information is associated with the identity of marketplace members, the most important being reputation. A reputation system must be in place to mitigate the effects of the information asymmetries that we have so far discussed. Part of the adverse selection problem revolves around uncertainty. The reality is that malicious actors exist, and they can be a threat to the health of the marketplace because they introduce the possibility of risk and loss for buyers and sellers. Buyers transacting with a malicious seller risk losing their investment by being sold junk data, and sellers transacting with a malicious buyer risk exploitation and having their privacy compromised. Reputation is a mechanism that limits the effects of malicious actors in the marketplace, and mitigates the effects of the adverse selection problem. 

As such, a reputation check within the identification function is monitored by the orchestrator. Reputation allows buyers and sellers to build rapport inside the marketplace as they complete transactions. The marketplace orchestrator monitors how the reputation of members fluctuates, and uses this information to advise members accordingly based on their potential transaction partner. The reputation of buyers and sellers can be harmed: if a seller exchanges data that is not as described, or corrupted in any way, their reputation will decrease flagging them as  a risky transaction partner. At the same time, buyers who engage in suspicious behaviour will also have their reputation decreased. Monitoring reputation allows that information to be later used as a passive form of risk modifier.

 A few options exist for implementing reputation, we simply describe it as a necessary mechanisms to mitigate marketplace risks. An example could be a dispute resolution system where sellers can raise claims of license violations to have a buyer's reputation lowered, or have them outrightly removed from the marketplace. Leaning too much on a description of the how the reputation mechanism is implemented is a slippery slope since it leads down to some of the more rampant and complex problems of digital platforms at large, such as fraud. It is essential though that reputation be included since it is a crucial mechanism that reinforces the risk mitigation component of our marketplace model. Our proposed implementation is to represent reputation using a graph, where nodes represent marketplace members, and edges represent completed transactions. In this representation, transaction edges are associated with a weight determined by the reputation of the transaction partner. This way, reputation is not only evaluated based on the number of completed transactions, but also by the aggregated reputation of all transaction partners. This kind of implementation is not heavy handed, is in line with our decentralized approach, and can be performed by distributed orchestrators.

Specification (2 in Figure~\ref{fig:transaction}) follows the identification function in the transaction flow. This function is common across both buyers and sellers, but the actual details of how it is performed differ between these actors. During specification, buyers determine what kind of data they are searching for, according to what access and use details. Additional specifications determined by the buyer can include the reputation of sellers listing the product, as well as the level of noise included in the data listed as a risk modifier. Sellers specifications are utilized in the product generation function further down the transaction flow. Sellers must specify the type of data they want to sell, which includes categorical details about the data such as qualities like range of time during which the data was collected, and quantity of data points in the listing. Additionally, the seller must specify their privacy preferences to help apply risk modifiers, as well as select a license that proposes terms of use and access that correspond to said preferences. 

Much like seller specifications inform the product generation function, buyer specifications inform the market search function (3 in Figure~\ref{fig:transaction}). Market search is undertaken exclusively by buyers. It is the point in the transaction flow where buyers are actively looking for product listings on the marketplace that match their specifications. Like most digital platforms, this process can be supported by additional search filters that help buyers find the exact kind of data they require. The counterpart to market search is the product generation function (4 in Figure~\ref{fig:transaction}). This point in the transaction flow is heavily informed by the seller's specifications, and it is supported by the orchestrator in terms of monitoring and logistics. There are four elements that make up the product generation phase: risk assessment, risk modification, licensing, and pricing. Risk assessment is performed according to the methodology outlined in our discussion of risk scores. Using the seller's specifications regarding the data provided, an assessment is made using our risk matrix to determine the magnitude of the risk represented by the sale of such data. In response to this risk assessment, risk modification can take place in the form of noise injection. The data can be distorted using noise as a privacy enhancing strategy. Once risk has been assessed, and possibly modified, the seller's specifications on terms of use and access are used to select a personal data license to formalize which personal rights are retained by the seller, and which are granted to the buyer. Finally, the pricing function is applied using input from the risk assessment and modification, as well as the licensing restrictions,  to generate a pricing recommendation that takes into account marketplace demand for the data already specified. The seller is entirely allowed to reject this pricing recommendation. 

The goal of providing such a recommendation is to inform the seller, and give them the aforementioned latitude regarding transaction control. We don't assume that sellers have an objective understanding of the value of personal data, hence the pricing recommendation using the pricing function serves as a guide for sellers. Seller's may overprice, or underprice their data as they wish, with positive or negative effects. They may not, however, reject the pricing modifications incurred by risk modification using noise. There is a discount applied to the product based on the level of distortion in the data that is essential to the risk modification mechanism. If sellers are allowed to sell noise-free data at the same price as distorted product counterparts, the entire risk modification mechanism is undermined as far as enhancing privacy via noise is concerned. Similarly, sellers may not reject the pricing modification incurred by having a low reputation in the marketplace due to the same reasoning.  

The product generation function is closely monitored by the orchestrator to ensure sellers do not undermine the effect of risk modifiers, as well as to ensure that sellers formalize their specifications via licensing. Once a price is set, the product generation function ends and the product itself is listed on the market . A market match takes place when a market search encounters a product listing, or a number of listings, that matches buyer specifications (5 in Figure~\ref{fig:transaction}). This brings the transaction flow to the first decision point: is the price acceptable to the buyer? If the price of the listing is acceptable, the transaction flow moves to the next function, otherwise the buyer may return to the listing match to consider other options, or may reconfigure their specifications and launch an entirely new market search. 

If the price is acceptable, the transaction flow continues to the subsample analysis function (6 in Figure~\ref{fig:transaction}).  The subsampling function serves to safeguard buyers from malicious sellers. It is a form of validation necessitated by the information asymmetry that is inherent to intangible products. We don't propose a specific subsampling technique as part of our model, but we do include it as a necessary mechanism to address the adverse selection and expropriation problems. As previously discussed, the adverse selection problem refers to an information asymmetry that's problematic for the sale of personal data: simply put, buyers cannot readily observe if the product they are interested in buying is legitimate or not from the product listing. Some kind of data validation is required to mitigate this asymmetry, which is the intention of introducing a subsampling mechanism. This leads us directly to the expropriation problem. Subsampling can expose sellers to privacy risks without compensation, as well as other kinds of market hazards like subsampling replication and other forms of fraud. For this reason, subsampling must be limited as a mechanism, and closely monitored by the orchestrator to ensure it is not used to manipulate the marketplace with malicious intentions. Once a buyer reaches the subsampling stage, they may accept or reject the subsample based on their original specifications for a market search. If the buyer rejects the subsample without any basis to do so, they risk different marketplace repercussions. Buyers that consistently demand subsamples, only to reject them will have their reputation lowered, and their ability to request subsamples suspended. 

If the buyer accepts the subsample, they are essentially recognizing the validity of the data in the product listing, as per their specifications, and the transaction flow can move to the final function: exchange (7 Figure~\ref{fig:transaction}). The exchange function can be implemented in different ways. In our model, it simply represents the exchange of data for financial compensation between buyer and seller. Depending on the amount of data, as well as the price to be paid, the details of the exchange may need to be adjusted. This is why the exchange function must be monitored by the orchestrator. Some exchanges may require an orchestrator to act as escrow between the transaction partners to ensure the exchange can be resolved adequately. In addition, the orchestrator is required to closely monitor the exchange because refusal of payment or refusal of data delivery at this point in the transaction flow must be reflected in the reputation of the transgressing transaction party, and penalties must be dealt accordingly. On the other hand, monitoring is also necessary so that the respective reputation of transaction partners can be updated in response to a successful exchange, thereby bringing the transaction flow to its end and completing our description of the transaction process.

\section{Conclusion}
We originally set out to answer the following question: what is the best model to use when dealing with the sale of personal data while keeping privacy intact? As a response, we analysed personal data in the abstract to distill the challenging qualities that had to be addressed by a model in which personal data can be transacted. In doing so, we outlined some key challenges. First, the challenges brought on by a personal data marketplace, like the adverse selection and expropriation problems. Second, the challenges brought on by personal data in the marketplace, such as rivalry and excludability. From these challenges we moved on to provide a definition of personal data in the marketplace context, as well as to describe its value. Our proposed marketplace model serves as a response to these challenges, as well as a blueprint for the implementation of a decentralized personal data marketplace that gives data generators the ability to protect their privacy and retain control of their data. 

As mentioned in our introduction, our proposed marketplace model is conceptual. It is not a directive on implementation. We recommend the use of some technologies, and discourage architectures that are not user centric or heavily centralized. The field of IoT and personal data has a bright future, and a long road ahead of it. Our proposed model can serve as a guide for further developments and architectures that deal with IoT and personal data, as well as a starting point for discussions on privacy centric personal data marketplaces.

\section*{Acknowledgment}

This work was funded by a grant of the National Science and Engineering
Research Council.

\ifCLASSOPTIONcaptionsoff
  \newpage
\fi

\end{document}